\documentclass[twocolumn,floatfix,pre,aps,showpacs]{revtex4}

\usepackage{graphicx}
\usepackage{amsmath}
\usepackage{epsfig}

\newcommand{\be}{\begin{equation}}
\newcommand{\ee}{\end{equation}}
\newcommand{\bear}{\begin{eqnarray}}
\newcommand{\eear}{\end{eqnarray}}
\newcommand{\bse}{\begin{subequations}}
\newcommand{\ese}{\end{subequations}}

\begin{document}

\title{Controlled Parameter Modulations in Secure Digital Signal Transmissions}

\author{P.~Palaniyandi}
\email{palani@cnld.bdu.ac.in}
\author{M.~Lakshmanan}
\email{lakshman@cnld.bdu.ac.in}
\affiliation{Centre for Nonlinear Dynamics, School of Physics,
Bharathidasan University, Tiruchirapalli 620 024, India}


\begin{abstract}
We propose a simple method for secure digital signal transmission by making 
some modifications in the single-step parameter modulation technique proposed
earlier into overcome certain inherent deficiencies.  In the modified method,
the parameter modulation is effectively regulated or controlled by the chaotic
signal obtained from the transmitting chaotic system so that it has the maximum
security.  Then, the same idea is also extended to the multistep parameter
modulation technique.  It is found that both the methods are secure against
ciphertext (return map) and plaintex attacks.  We have illustrated these methods
by means of the Lorenz system.
\end{abstract}


\maketitle

\section{Introduction}    
Sometime ago, Pecora and Carroll have shown that two identical chaotic systems
(namely, drive and response systems) can be synchronized by an appropriate
coupling between them [Pecora \& Carroll, 1990; Pecora \& Carroll, 1991]. This
idea immediately motivated several researcher to use chaotic signal for secure
communication purpose with the belief that it will be difficult for an intruder
to decipher the transmitted message without the knowledge of the transmitting
chaotic system, due to its extremely high sensitiveness to the initial
conditions and parameter mismatches.  As a result, a number of methods have been
proposed for secure communication and cryptography [Hayes et al,  1993; Cuomo \&
Oppenheim, 1993; Murali  \& Lakshmanan, 1993a;  Murali \& Lakshmanan, 1993b;
Kocarev \& Parlitz,1995; Lakshmanan \& Murali, 1996; Zhang et al, 1998;
Baptista, 1998;  He \& Vaidya,1998;  Wong et al, 2003; Kocarev et al, 2004;
Bowong, 2004; Hua et al, 2005; Chee \& Xu 2006]  using chaotic signals. In
particular, Cuomo and Oppenheim [Cuomo \& Oppenheim, 1993] have suggested a very
simple method for secure digital signal transmission using the property of the
coupled chaotic systems that a small difference between the corresponding
parameters in the drive and response systems will cause synchronization
frustration in their dynamical variables. Here, the driving signal which drives
the response or receiver system is used as a carrier signal in the digital
signal transmission. During transmission, the digital message is imposed on this
carrier signal through parameter variations. So, one may call the driving signal
which implicitly bears the digital message as modulated driving signal and the
process as parameter modulation.  However, P\'erez and Cerdeira [P\'erez \&
Cerdeira, 1995] have shown that it is possible to reconstruct this masked
message by an eavesdropper from a simple return map formed by the extrema of the
modulated driving signal, even without any knowledge about the chaotic systems
(transmitter and receiver).  To overcome the return map attack, several methods
have been suggested by many authors [Murali  \& Lakshmanan, 1998; Mensour \&
Longtin, 1998; Minai \& Pandian, 1998; Palaniyandi \& Lakshmanan, 2001]: 
Private communication using compound chaotic signal technique, using
delay-differential equations technique, communication through noise, multistep
parameter modulation and so on.  However, most of these methods are found to be
very difficult to implement, and some of them have been shown to be not so
secure as expected.   For example, we have proposed a method called multistep
parameter modulation to complicate the patterns in the return map so that the
reconstruction of the message by an eavesdropper is almost impossible
[Palaniyandi \& Lakshmanan, 2001].  However,  Li et al [Li et al, 2006] have
very recently pointed out that the multistep parameter modulation suggested by
us may not be so secure as was expected for smaller modulation steps ($n$).
These authors have also pointed out that the multistep parameter modulation
technique is secure only if $n>50$, instead of $n>17$ expected in our analysis.
In order to remove such difficulties, in this Letter,  we introduce a new
technique called controlled parameter modulation in which the modulation is
effectively directed or regulated by the chaotic signal obtained from one of the
dynamical quantities of the transmitter, instead of external predetermined
regulation where a particular value of modulation parameter is preassigned for
transmitting `$0$' or `$1$' bit in the digital message.

The Letter is organized as follows.  In Secs.~\ref{ss_pm}\&\ref{ms_pm}, we
briefly outline the single-step and multistep parameter modulation techniques
and the possible cryptographic attacks on these methods.  To overcome the
drawbacks of these techniques, a controlled parameter modulation technique is
introduced in Sec.~\ref{cont_pm}.  The controlled single-step parameter
modulation technique is illustrated and its security against various
cryptographic attacks is given Sec.~\ref{cont_ss_pm}.  The same analysis is done
for the controlled multistep parameter modulation technique in
Sec.~\ref{cont_ms_pm}.  Finally, in Sec.~\ref{con}, we summarize the results of
our analysis.

\section{Single-step Parameter Modulation}
\label{ss_pm}
Let us now outline the method of digital signal transmission by single-step
parameter modulation proposed by Cuomo and Oppenheim, illustrated by means of
the Lorenz system [Cuomo \& Oppenheim, 1993].  In this case, the chaotic signal
is produced at the transmitting end by  
\bse
\label{transmitter}
\bear
\dot{x}_s &=& \sigma (y_s-x_s), \\ 
\dot{y}_s &=& rx_s-y_s-x_sz_s, \\ 
\dot{z}_s &=& x_sy_s-bz_s,  
\eear
\ese
where the parameters $\sigma$ and $r$ take the values $16.0$ and $45.6$,
respectively. The remaining parameter $b$ is chosen for the purpose of
modulation, which is assigned (for illustration) either the value $4.0$ or $4.4$
depending upon the nature of the digital information to be transmitted. At the
receiving end, the chaotic signals are generated from the system 
\bse
\label{receiver_x}
\bear
\dot{x}_r &=& \sigma (y_r-x_r), \\
\dot{y}_r &=& rx_s-y_r-x_sz_r, \\ 
\dot{z}_r &=& x_sy_r-bz_r, 
\eear
\ese
where $\sigma=16.0$, $r=45.6$ and $b=4.0$.  Note that the receiver is driven by
$x_s$ and the value of modulation parameter ($b$) has been now fixed in the
receiver.  The coupling of this type was introduced by Pecora and Carroll
[Pecora \& Carroll, 1990] for achieving synchronization between two identical
chaotic systems.

The transmission of digital information is done as follows. The value of
modulation parameter ($b$) is switched between $4.0$ and $4.4$ in the
transmitter according to the nature of the digital message. Due to this
switching in the values of modulation parameter $b$, the receiver is driven
either by $x_s$ corresponding to the set of values of the parameters $16.0$,
$45.6$ and $4.0$, for $\sigma$, $r$ and $b$, respectively, if the transmitted
message bit is `$0$', or by $x_s$ corresponding to another set of parameter
values  $16.0$, $45.6$ and $4.4$, for $\sigma$, $r$ and $b$, respectively, if
the transmitted message bit is `$1$'.  Note that the transmitting and receiving
chaotic systems have identical sets of values of the parameters while a binary
state `$0$' is transmitted, and there is a variation in the corresponding values
of b in these systems while the other binary state `1' is transmitted.  As a
result, the receiver will synchronize with the transmitter when the message bit
transmitted is `$0$' and asynchronization takes place if the transmitted message
bit is `$1$'.  Then the transmitted digital bit is reconstructed at the
receiver   using the synchronization error power $(x_r-x_s)^2$, since it is
negligible when the transmitter and receiver systems are synchronized, and it
has some finite value when they are not synchronized.  We can call the above
procedure as a single-step parameter modulation, since the modulation parameter
$b$ can have only one value for each state of the binary message (that is,  in
this method of  transmission, $b$ can take only a single value, $4.4$ for any
high state (`$1$') in the digital message and $4.0$ for the low state (`$0$') of
digital message at the transmitter).

\subsection{Return Map Attack} 
\label{ret_map_attack} 
\begin{figure} 
\begin{center} 
\epsfig{figure=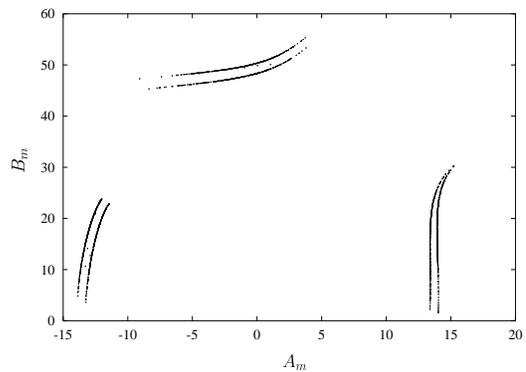,width=0.8\columnwidth} 
\end{center} 
\caption{Return map between $A_m$ and $B_m$ obtained from the modulated driving
signal $x_s$ of the Lorenz system in single-step parameter modulation.}
\label{retmapx_ss_lor}
\end{figure}

Eventhough the single-step parameter modulation technique is very simple and
found to be secure enough against many possible security attacks proposed by
Short [Short, 1994; Short, 1996; Short, 1997; Short, 1998],  P\'erez and
Cerdeira have shown that the message can be extracted or unmasked from the
return map constructed from the modulated drive signal, without any receiver
circuit [P\'erez \& Cerdeira, 1995]. In their work, the maxima $X_m$ and minima
$Y_m$ are collected from the modulated driving signal ($x_s$) and two new
variables $A_m=(X_m+Y_m)/2$ and $B_m=X_m-Y_m$ are defined. Then a return map is
plotted between $A_m$ and $B_m$ as shown in Fig.~\ref{retmapx_ss_lor}. Note that
there are 3 segments in the attractor of the return map, each one further splits
into two strips. It is then obvious to assume that the split in the return map
attractor is due to the change in the values of the parameter $b$ at the
transmitter between $4.0$ and $4.4$ (hence it may be assumed that one strip in
each segment corresponds to the high state and the other corresponds to the low
state of the digital message).  From this return map one can easily unmask the
message by noting which points $(A_m$, $B_m)$, fall on which strips in each
segment at various instants of time.  If we assume that $6$ strips are
independent of each other, then the attack complexity is $8$ [P\'erez \&
Cerdeira, 1995; Palaniyandi \& Lakshmanan, 2001].
 
\section{Multistep Parameter Modulation}
\label{ms_pm}
In order to complicate the patterns in the return map, we have proposed a
multistep parameter modulation where `$1$' bit is transmitted using a set of $n$
predetermined values of the parameter $b$ and `$0$' bit is transmitted by making
use of another set of $n$ predetermined values of $b$ ($n>1$ and sufficiently
large) [Palaniyandi \& Lakshmanan, 2001]. At the receiver end, we use $n$
receiver subsystems, each one with a different value of $b$.  The set of values
of $b$ assigned at the receiver is nothing but the set of $n$ values of $b$
which are used in the transmitter for modulating a high state of the digital
message.  We call this number $n$ as the step of the modulation.  

Let us now describe this method for a modulation step $5$. It can be easily
verified that the Lorenz system~(\ref{transmitter}) exhibits chaotic behavior
when the parameter  $b$ takes any value between  $1.5$ and $6.8$ while the other
parameters are fixed at $\sigma=16.0$ and $r=45.6$.  As an illustration, the
parameter $b$ in the transmitting chaotic system is allowed to take any one of
the five values $3.1, 3.3, 3.5, 3.7$, and $3.9$, for transmitting a high state
(`$1$'), while it can have any one of the other five values $3.2, 3.4, 3.6,
3.8$, and $4.0$, for transmitting a low state (`$0$').  In the receiver part, we
use $5$ subsystems with the modulation parameter $b$ fixed at $3.1, 3.3, 3.5,
3.7,$ and $3.9$.

The method works as follows. Suppose we have a high state in the digital
message, to start with. Then, the receiver is driven by the modulated driving
signal obtained for the modulation parameter $b=3.1$.  For the next high state
in the message, the modulation is done with $b$ = $3.3$ and this process
continues upto the value $3.9$.  Then the value of $b$ is reset to $3.1$.  The
same procedure is followed for transmitting the low state of the message but
with the modulation parameter ($b$) taking the values $3.2, 3.4, 3.6, 3.8,$ and
$4.0$ in that order.  In the receiver end,  all the subsystems are driven by the
modulated driving signal.  If the transmitted message is in the high state, then
one of the subsystems in the receiver will be synchronized with the transmitter.
If none of the subsystems is synchronized, then it is taken that the message
transmitted is in the low state.   
\begin{figure} 
\begin{center} 
\epsfig{figure=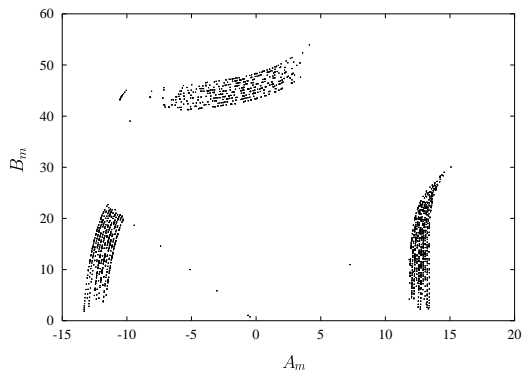,width=0.8\columnwidth} 
\end{center} 
\caption{Return map between $A_m$ and $B_m$ calculated from the modulated
driving signal $x_s$ of the Lorenz system in multistep parameter modulation
$(n=5)$.}
\label{retmapx_ms_lor}
\end{figure}

\subsection{Return Map/Ciphertext Attack} 
The return map constructed from the modulated  driving signals used in multistep
parameter modulation technique is shown in Fig.~\ref{retmapx_ms_lor}. Now, we
have 10 strips in each segment.  This corresponds to $2 \times n$ strips, where
$n$ is the step of the parameter modulation.  Because of this increase in number
of strips in each segment, it was claimed in our earlier paper [Palaniyandi \&
Lakshmanan, 2001] that it would be difficult to unmask the message from the
return map.  However,  Li et al [Li et al, 2006] have very recently pointed out
that the multistep parameter modulation suggested by us may not be so secure as
was expected for smaller modulation steps.  These authors have also noted that
there exists a deterministic relationship between the positions of the $2n$
strips of a segment and the $2n$ different values of the modulation parameter
`$b$', and this relationship will reduce the attack complexity of the return
map.  This can be easily viewed in the single-step parameter modulation (from
Fig.~{\ref{retmapx_ss_lor}}) that the strips corresponding to $b=4.0$ are closer
to the origin in all the three segments and the strips corresponding to $b=4.4$
are away from the origin.  This means that there exists only two different
possibilities of assigning $0/1$ bit to the strips in all the segments instead
of eight: Assign $0$-bit ($1$-bit) to the strips closer to the origin and
$1$-bit ($0$-bit) to the strips away from the origin.  The above analysis can be
possibly generalized to the multistep parameter modulation and hence the task of
assigning $0/1$ bit to $6n$ strips ($2n$ strips for each segment) in the return
map is reduced to an another task of assigning $0/1$ bit to $2n$ strips or $2n$
different vales of $b$.  Thus, the attack complexity of the return map obtained
in the case of multistep parameter modulation becomes $2^{2n}-2$, or
approximately, $2^{2n}$ instead of $2^{6n}$.  Due to today's advancement in the
computer technology,  a practically secure cryptosystem is acceptable only if it
has an attack complexity greater than $2^{100}$.  As a result, the multistep
parameter modulation technique is secure only if $n>50$, instead of our earlier
envisaged modulation step $n\approx17$.  Eventhough it is practically possible
to implement the multistep parameter modulation with modulation step greater
than 50, it will become more expensive. For this reason, we look for an
appropriate improvement in the parameter modulation technique and is given in
the following section.  

\subsection{Plaintext Attack} 
Now let us consider the security of the multistep parameter modulation against
known-plaintext or chosen-plaintext attacks.  In known/chosen-plaintext attacks,
it is obvious that the knowledge about some plaintexts means the knowledge about
some bit assignment of the $6n$ strips in the return map: When the message
transmitted is known to be $0$-bit ($1$-bit), one immediately concludes that the
strip on which the point $(A_m,\,B_m)$ lies corresponds to the $0$-bit
($1$-bit).  Once $n$ $0$-bits ($1$-bits) have been assigned to $n$ different
strips, that is, to $n$ different values of $b$, the attacker can directly
assign $1$-bits ($0$-bits) to all the remaining strips, that is, to the
undetermined values of $b$, so as to complete the attack.  

\section{Controlled Parameter Modulation}  
\label{cont_pm}
In order to overcome various possible cryptographic attacks [Short, 1994; Short,
1996; Short, 1997; Short, 1998; P\'erez \& Cerdeira, 1995; Li et al, 2006], we
have made a slight but effective modification in the parameter modulation
technique as explained below. We wish to note that the chaotic drive system
possesses not only the chaotic signal $x_s$, but also the other two chaotic
signals $y_s$ and $z_s$. In general, there exists no simple relation between
these signals.  So, one can make effective use of the remaining signals (other
than that of $x_s$) in the process of switching of the modulation parameter so
that any particular strip in the return map will represent both the `0' and  `1'
states.  That is, the switching is carried out as a function of the
instantaneous values of $y_s$ and $z_s$ at some time. If we incorporate these
changes, then the switching in the values of modulation parameter will depend on
the dynamics of the transmitting chaotic system, in addition to the nature of
the digital message to be transmitted.  That is, the modulation is now
controlled or influenced by the state values of the variables $y_s$ and $z_s$ of
the transmitting chaotic system.  We may call this modified method as the signal
controlled parameter modulation, or simply, controlled parameter modulation.
This idea can be implemented both in the single-step parameter modulation and in
the multistep parameter modulation techniques as illustrated below.

\section{Controlled Single-step Parameter Modulation} 
\label{cont_ss_pm}

In the single-step parameter modulation described in Sec.~\ref{ss_pm}, the value
of $b$ is switched between $4.0$ and $4.4$ in the transmitter for imposing the
digital information `$0$' and `$1$' bits, respectively, on the $x_s$ signal.
Hence the assignment of the values to the parameter $b$ depends only on the
nature of the digital message. But in the controlled single-step parameter
modulation, we introduce an additional condition (any one of the conditions
listed below) while assigning the above values to the parameter `$b$' during the
transmission process.  As a simple example, we note the value of the chaotic
signal corresponding to the $y$ variable of the transmitting chaotic system at
the time of switching of binary states (that is, from `0' to `1' or vice versa)
if adjacent bits are different, or at the time of transmitting the edge of the
width of the first bit if adjacent bits are identical.  The value of $y$ at this
instant is represented by $y_{swt}$ and it is used as the additional
condition/parameter for determining the values of the parameter $b$.  That is,
$y_{swt}$ is now used to control the single-step parameter modulation. A simple
way of using $y_{swt}$ to control the single-step parameter modulation is the
following: If $y_{swt}>0$,  the modulation parameter $b$ takes the value $4.4$
while transmitting `$1$' bits, and $4.0$ for transmitting `$0$' bits.  On the
other hand, if $y_{swt}<0$, the modulation parameter $b$ is assigned to have the
values in the opposite manner, that is, $b=4.0$ for sending `$1$' bits, and
$b=4.4$ for sending `$0$' bits.  Note that the binary bit 0/1 is now modulated
using both the values of the modulation parameter $b$, namely, $4.0$ and $4.4$. 
At the receiver end, we incorporate two subsystems driven by the modulated
driving signal $x_s$, one with the parameters $\sigma=16.0$, $r=45.6$ and
$b=4.0$, and the other with $\sigma=16.0$, $r=45.6$ and $b=4.4$. Since only
these two sets of values of the parameters are used in the transmitter for
modulation purpose, at the time of switching of binary states, one of the
subsystems is certainly synchronized with the transmitting chaotic system
whatever be the message transmitted. Simultaneously, it is possible to obtain
$y_{swt}$ from the synchronized receiver subsystem since $y_r=y_s$ after
transient time.  Since the receiver knows the rule of control on the parameter
modulation before hand, the message can be constructed from the modulated
driving signal $x_s$ by finding which of the subsystems in the receiver is
synchronized with the transmitting chaotic system once $y_{swt}$ is obtained. 
Note that the signal $y_s$ is not transmitted to the receiver and so it is not
possible to obtain $y_{swt}$ by the intruder to decipher $x_s$ directly.  It is
important to note that the above is only an illustrative case and one can
consider even more complicated ways to control the parameter modulation through
various means: 

\begin{enumerate} 
\item  
By incorporating conditions on a complicated function of $y_s$ rather than on
$y_s$ itself.
\item 
by introducing conditions on $y_{s}$ and $z_{s}$ considered together (as
$y_{swt}$ \& $z_{swt}$).
\item
by introducing conditions on a complicated functions of $y_{s}$ and $z_{s}$.
\end{enumerate}  

and so on.

\subsection{Return Map/Ciphertext Attack}   
\begin{figure} 
\begin{center} 
\epsfig{figure=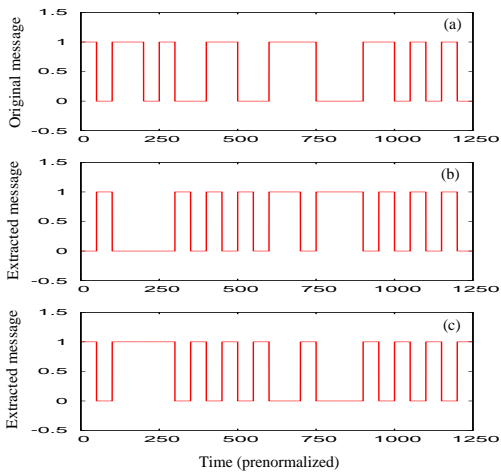, width=0.8\columnwidth} 
\end{center} 
\caption{Messages extracted from the return map in controlled single-step
parameter modulation: (a) Original message transmitted, (b) message obtained by
assigning strips closer to the origin correspond to the bit `0' and the
remaining strips correspond to `1' bit, and (c) message obtained by assigning
strips closer to the origin correspond to the bit `1' and the remaining strips
correspond to `0' bit.}
\label{mesg_ext}
\end{figure}

In this method, the return map constructed from the modulated driving signal is
identical with the return map in Fig.~\ref{retmapx_ss_lor}, since we have again
used (as in the case of single-step parameter modulation) the values $4.0$ and
$4.4$ for $b$ while modulating the digital message in the transmitter. 
Eventhough one can assume different strips correspond to the different values of
$b$, it is now uncertain to assign a binary bit $0/1$ to a particular strip in
the return map.  This uncertainty is due to the fact that both the values of
$b$, namely, $4.0$ and $4.4$ have been used in transmitting each state of the
binary bits, either `0' or `1' at various intervals of time. As a result, the
points ($A_m,\,B_m$) calculated from the modulated driving signal fall on both
the strips in all the segments of the return map while transmitting either the
binary state `0' or the binary state `1'.  Thus, it is not possible to extract
the original message from the return map unless one knows the rules by which the
values of $b$ are switched for transmitting the digital information.  As in the
case of single-step parameter modulation, if one assumes that the strips closer
to the origin corresponds to `0' bit and the strips away from the origin
corresponds to `1' bit or vice versa [Li et al, 2006], then it is possible to
obtain two different messages as shown in Fig.~\ref{mesg_ext}.  From this
figure, it is observed that the probability of extracting the original message
from the return map is practically negligible.  Hence, the controlled
single-step parameter modulation is secure against the return map attack.  
    
\subsection{Plaintext Attack}
If there is a chance of finding the values of $b$ which have been used in the
transmitting chaotic system from the return map, an intruder can construct the
subsystems identical to the receiver subsystems and they can be driven by the
modulated driving signal $x_s$.  Eventhough it is almost impossible, still there
is a chance to find the rules by which the parameter modulation is controlled
with the help of known plaintexts and the synchronization of intruder's
subsystems with transmitting chaotic system.  However, even this possibility can
be eliminated if we use complicated functions or time-varying functions of $y_s$
and $z_s$ (collected at various instants of time) for controlling the parameter
modulation. So, the controlled single-step parameter modulation appears to be
reasonably secure even against the plaintext attack.

\section{Controlled Multistep Parameter Modulation}
\label{cont_ms_pm}    
To illustrate the controlled multistep parameter modulation, we again consider
the Lorenz system with a set of values of parameters used in Sec.~\ref{ms_pm}
and a modulation step of $n=5$. Unlike the case of standard multistep parameter
modulation, now the digital message `$1$' is transmitted with one of the values
of the modulation parameter $b$ in the set \{$3.1,3.3,3.5,3.7,3.9$\}, and `$0$'
is transmitted with one of the values of $b$ from the set
\{$3.2,3.4,3.6,3.8,4.0$\}, when $y_{swt}$ is positive. On the other hand, if
$y_{swt}$ is negative, the set of values \{$3.2,3.4,3.6,3.8,4.0$\} is used for
the modulation while sending  `$1$' bits in the digital message and the set of
values \{$3.1,3.3,3.5,3.7,3.9$\} is used in the modulation for sending `$0$'
bits in the message.  It may be noted that all the $10$ values of $b$ have been
used for transmitting each state, that is, either `$1$' or `$0$' of the digital
message.

The transmission is done as follows.  Suppose we have the bit `$1$' in the
digital message to start with, then it is transmitted by assigning 3.1 to the
modulation parameter $b$.  If the next bit is `$1$', it is then modulated using
$3.3$ if $y_{swt}$ is greater than zero, or using $3.2$ if $y_{swt}$ is less
than zero.  On the other hand, if the second bit is `$0$', then the modulation
is done using $3.2$ if $y_{swt}$ is greater than zero, or it is done using
$3.3$, if $y_{swt}$ is less than zero.  This process continues upto $3.9$ and
$4.0$ in each set of values of the modulation parameter $b$ and then it is reset
to $3.1$ and $3.2$ in the respective sets.  Thus, we have used all the values
($3.1,3.2,\dots,4.0$) of $b$ in transmitting each state of the binary message. 

In the receiver end, there are $10$ subsystems driven by the modulated driving
signal $x_s$, each one with different values of $b$ from the set \{$3.1, 3.2,
3.3, 3.4, 3.5, 3.6, 3.7, 3.8, 3.9, 4.0$\}. Since one of these values is used at
the transmitter end while transmitting the digital message, synchronization
between the transmitting chaotic system and one of the receiver subsystems is
always achieved.  Hence, it is possible to obtain $y_{swt}$ in the receiver end,
from the synchronized subsystem where $y_r=y_s$ at any instant of time after
transient die down.  As the nature of control on the parameter modulation is
known at the receiver end, the message can be constructed from the modulated
driving signal $x_s$ by finding which of the subsystems in the receiver is
synchronized with the transmitting chaotic system, once $y_{swt}$ is obtained.

\subsection{Return Map Attack}   
Now, the return map constructed from the modulated driving signal is identical
with the return map in Fig.~\ref{retmapx_ms_lor}, since we have again used (as
in the case of multistep parameter modulation) the values from the \{$3.1, 3.2,
3.3, 3.4, 3.5, 3.6, 3.7, 3.8, 3.9, 4.0$\} for $b$ while modulating the digital
message in the transmitter.  Eventhough one can assume different strips
correspond to the different values of $b$, it is not possible to assign a binary
bit $0/1$ to a particular strip in the return map.  Because, now all the $10$
values of $b$ have been used in transmitting each state of the binary bits,
either `0' or `1' at various interval of time. As a result, the points
($A_m,\,B_m$) calculated from the modulated driving signal fall on all the $10$
strips of a segment in the return map while transmitting either the binary state
`0' or the binary state `1'.  Thus, it is not possible to extract the original
message from the return map unless one knows the means by which the values of
$b$ are switched for transmitting the digital information. Suppose the attacker
randomly assume the binary state 0/1 to each strip in the return map, then it is
possible to construct $2^{2n}-2$ different messages, that is, $1022$ different
messages can be constructed for $n=5$.  In general, none of the messages will
resemble the original message.  Thus, there is practically negligible chance for
obtaining the exact message from the return map provided the message is long
enough.

\subsection{Plaintext Attack} 
As in the case of the controlled single-step parameter modulation, controlled
multistep parameter modulation is secure against the plaintext attack if one
uses the functions or time-varying functions of $y_s$ and $z_s$ (collected at
various instants of time) for controlling the parameter modulation.

\section{Conclusion}  
\label{con} 
We have proposed a controlled parameter modulation technique (applicable to both
the single-step and multistep parameter modulations) wherein parameter
modulation is controlled by a chaotic signal (other than the driving signal)
produced in the transmitter. These modified methods have been illustrated for
the Lorenz system. Also, it has been shown that the controlled single-step
parameter modulation and controlled multistep parameter modulation techniques
are secure enough against both the ciphertext and plaintext attacks.  

\acknowledgments
This work has been supported by the National Board for Higher Mathematics,
Department of Atomic Energy, Government of India and the Department of Science
and Technology, Government of India through research projects.

\section*{REFERENCES} 

\begin{description}

\item  Baptista, M. [1998] ``Cryptography with chaos" Phys. Lett. {\bf A240},
50-54. 

\item Bowong, S. [2004] ``Stability analysis for the synchronization of chaotic
systems with different order: application to secure communications" Phys. Lett.
{\bf A326}, 102-113. 

\item Chee, C. Y. \& Xu, D. [2006] ``Chaotic encryption using discrete-time
synchronous chaos''   Phys. Lett. {\bf A348}, 284-292.

\item Chua, L.O., Kocarev, Lj., Eckert, K. \& Itoh, M. [1992] `` Experimental
chaos synchronization in Chua's circuit" Int. J. Bifurcation and Chaos {\bf 2},
705-708.

\item Cuomo, K.M. \& Oppenheim, A. V. [1993] ``Circuit implementation of
synchronized chaos with application to communication" Phys. Rev. Lett.  {\bf
71}, 65-68.

\item Hayes, S., Grebogi, C. \& Ott, E. [1993] ``Communicating with chaos"
Phys.  Rev. Lett. {\bf 70}, 3031-3035.

\item He, R. \& Vaidya, P. G. [1998]  ``Implementation of chaotic cryptography
with chaotic synchronization" Phys. Rev. {\bf E57}, 1532-1535.

\item Hua, C. Yang, B. Ouyang, G \& Guan, X. [2005] ``A new chaotic secure
communication scheme" Phys. Lett. {\bf A342}, 305-308.

\item Kocarev, L. \& Parlitz, U. [1995] ``General approach for chaotic
synchronization with applications to communication"  Phys. Rev. Lett. {\bf 74},
5028-5031.

\item  Kocarev, L. Sterjev, M. Fekete, A \& Vattay, G. [2004] ``Public-key
encryption with chaos" Chaos {\bf 14}, 1078-1082.

\item Lakshmanan, M. \& Murali, K. [1996] Chaos in Nonlinear Oscillators:
Controlling and Synchronization (World Scientific, Singapore).

\item  Li, S. \'Alvarez, G. \& Chen, G. [2006] ``Return-Map Cryptanalysis
Revisited" Int. J. of Bifurcation and Chaos {\bf 16}.

\item Madan, R. N., [1993] Chua's circuit: A Paradigm for Chaos (World
Scientific, Singapore).

\item Mensour, B. \& Longtin, A. [1998] ``Synchronization of delay-differential
equations with application to private communication" Phys. Lett. {\bf A244},
59-70.

\item Minai, A. A. \& Pandian, T. D. [1998] ``Communicating with noise:  How
chaos and noise combine to generate secure encryption keys" Chaos {\bf 8},
621-628.

\item Murali, K. \& Lakshmanan, M. [1993a] ``Transmission of signals by
synchronization in a chaotic van der pol-duffing oscillator" Phys. Rev. {\bf
E48}, R1624-1626  

\item Murali, K. \& Lakshmanan, M. [1993b] ``Synchroning chaos in driven chua's
circuit" Int. J. Bifurcations Chaos {\bf 3}, 1057-1066.

\item Murali, K. \& Lakshmanan, M. [1998] ``Secure communication using a
compound signal from generalized synchronizable chaotic systems" Phys.  Lett.
{\bf A241}, 303-310.

\item Palaniyandi, P \& Lakshmanan, L. [2001] ``Secure digital signal
transmission by multistep parameter modulation and alternative driving of
transmitter variables"  Int. J. of Bifurcation and Chaos {\bf 7}, 2031-2036.

\item Pecora, L. M. \& Carroll, T. L. [1990] ``Synchronization in chaotic
systems" Phys. Rev. Lett. {\bf 64}, 821-823 (1990);

\item Pecora, L. M. \& Carroll, T. L. [1991] ``Driving systems with chaotic
signals" Phys. Rev. {\bf A44}, 2374-2383.

\item P\'erez, G. \& Cerdeira, H.A. [1995] ``Extracting messages masked by
chaos" Phys. Rev. Lett. {\bf 74}, 1970-1973.

\item  Short, K. M. [1994] ``Steps toward unmasking secure communications" Int.
J. of Bifurcation and Chaos {\bf 4}, 959-977.

\item  Short, K. M. [1996] ``Unmasking a modulated chaotic communications
scheme" Int. J. of Bifurcation and Chaos {\bf 6}, 367-375.

\item  Short, K. M. [1997] ``Signal extraction from chaotic communications"
Int. J. of Bifurcation and Chaos  {\bf 7}, 1579-1597.

\item  Short, K. M. \& Parker, A. T.  [1998] ``Unmasking a hyperchaotic
communication scheme"  Phys. Rev. {\bf E58}, 1159-1162.

\item Wong, K. Ho, S. \& Yung, C.  [2003] ``A chaotic cryptography scheme for
generating short ciphertext"  Phys. Lett. {\bf A310}, 67-73.

\item Zhang, Y., Dai, M., Hua, Y., Ni, W, \& Du, G. [1998] ``Digital
communication by active-passive-decomposition synchronization in hyperchaotic
systems" Phys. Rev. {\bf E58}, 3022-3027.

\end{description}

\end{document}